\documentclass{article}

\usepackage{subcaption}

\usepackage{siunitx}
\usepackage{PRIMEarxiv}
\usepackage{capt-of}
\usepackage{siunitx}
\usepackage[utf8]{inputenc} 
\usepackage[T1]{fontenc}    
\usepackage{hyperref}       
\usepackage{url}            
\usepackage{booktabs}       
\usepackage{amsfonts}       
\usepackage{nicefrac}       
\usepackage{microtype}      
\usepackage{lipsum}
\usepackage{amsmath}
\usepackage{fancyhdr}       
\usepackage{graphicx}       
\usepackage{color}
\graphicspath{{media/}}     

\usepackage{setspace}

\title{Modeling Inverse Demand Function with Explainable Dual Neural Networks
}

\author{
  Zhiyu Cao \thanks{Equal Contribution.}\\
  Stevens Institute of Technology \\
  Hoboken, NJ, USA \\
  \texttt{zcao17@stevens.edu} \\
  \And
  Zihan Chen \footnotemark[1] \\
  Stevens Institute of Technology \\
  Hoboken, NJ, USA \\
  \texttt{zchen61@stevens.edu} \\
  \And
  Prerna Mishra \\
  Georgia State University \\
  Atlanta, GA, USA \\
  \texttt{pmishra4@gsu.edu} \\
  \And
  Hamed Amini \thanks{I would like to thank Alireza Aghasi for helpful discussions.} \\
  University of Florida \\
  Gainesville, FL, USA \\
  \texttt{aminil@ufl.edu} \\
  \And
  Zachary Feinstein \\
  Stevens Institute of Technology \\
  Hoboken, NJ, USA \\
  \texttt{zfeinste@stevens.edu} \\
}

\begin{document}
\maketitle

\begin{abstract}
Financial contagion has been widely recognized as a fundamental risk to the financial system. Particularly potent is price-mediated contagion, wherein forced liquidations by firms depress asset prices and propagate financial stress, enabling crises to proliferate across a broad spectrum of seemingly unrelated entities. Price impacts are currently modeled via \emph{exogenous} inverse demand functions. However, in real-world scenarios, only the initial shocks and the final equilibrium asset prices are typically observable, leaving actual asset liquidations largely obscured. This missing data presents significant limitations to calibrating the existing models. To address these challenges, we introduce a novel dual neural network structure that operates in two sequential stages: the first neural network maps initial shocks to predicted asset liquidations, and the second network utilizes these liquidations to derive resultant equilibrium prices. This data-driven approach can capture both linear and non-linear forms without pre-specifying an analytical structure; furthermore, it functions effectively even in the absence of observable liquidation data. Experiments with simulated datasets demonstrate that our model can accurately predict equilibrium asset prices based solely on initial shocks, while revealing a strong alignment between predicted and true liquidations. Our explainable framework contributes to the understanding and modeling of price-mediated contagion and provides valuable insights for financial authorities to construct effective stress tests and regulatory policies.
\end{abstract}



\section{Introduction}

The interconnections among financial institutions can propagate and amplify shocks during financial crises \cite{braouezec_strategic_2019, glasserman_how_2015}, the risks of which are often unexpected \emph{a priori}. One striking instance is the 2007 subprime crisis, where subprime mortgage backed securities -- initially perceived as a minor asset class -- unleashed a wave of catastrophic economic losses and incited a global recession \cite{brunnermeier_deciphering_2009}. This progression, wherein exogenous shocks snowball into substantial losses, is conceptualized as contagion effects. Among the various contagion types, price-mediated contagion is notably potent: When encountering external shocks, firms may be forced to liquidate portions of their holdings, which depresses the asset's price and imposes financial stress on all entities possessing the same asset. As these distressed institutions strive to stabilize their balance sheets by liquidating other assets, this behavior becomes self-reinforcing throughout broader financial markets \cite{shleifer_fire_2011}. Considering price-mediated contagion permits the crisis to spread through firms that may not even have direct contractual relations, it can ripple across a wide range of seemingly unrelated entities and magnify the overall financial damage \cite{greenwood_vulnerable_2015,calimani2022simulating}. Thus, the development of models that accurately depict how adverse shocks spur firms to liquidate assets, and how such actions subsequently affect prices, is of paramount importance in mitigating systemic risk within financial markets.

To measure the impact of asset sales on their prices, existing studies propose ``inverse demand functions'' to mathematically model the relationship between trading activities and resultant prices. However, when examining price-mediated contagion via these functions, the literature presents two significant gaps. First, in real-world scenarios, typically only the initial shocks and the final equilibrium asset prices are observable. The realized asset liquidation or portfolio rebalancing actions undertaken by financial institutions tend to remain obscured, thereby complicating the direct modeling process from liquidation to price \cite{feinstein_effects_2017}. Second, current literature utilizes exogenous forms, primarily either a linear or exponential form, to model the price impacts \cite{cifuentes_liquidity_2005, greenwood_vulnerable_2015}. Despite their ease of interpretation, these forms fail to deliver a comprehensive financial interpretation or a compelling economic rationale in practice \cite{bichuch_endogenous_2022}.

Aiming to address these gaps, we introduce a novel dual network structure that models the inverse demand function in two stages: (i) The initial shocks are mapped into predicted equilibrium ``liquidations'' through the first neural network, and (ii) these predicted liquidations are then input into the second network to generate resultant asset prices. This methodology offers two primary advantages: First, rather than pre-specifying an analytical form, we employ deep neural networks to model the inverse demand function, which enables us to capture both linear and non-linear constructions based on realized data. Second, our framework is capable of modeling the inverse demand function even in the absence of observable ground truth liquidation data. 

Moreover, our work also contributes to the explainable machine learning (XAI) literature in the financial economics realm. Although the advent of machine learning (ML) models has led to substantial transformations in the finance sector, the inherent lack of transparency and explainability in these models presents significant challenges for users and regulators in establishing trust \cite{gomber_fintech_2018, misheva_explainable_2021}. In contrast to existing XAI approaches that explain the model \emph{ex post} \cite{du_techniques_2019}, our framework endows each model component with explicit financial significance. Specifically, we define the hidden liquidation as a learnable intermediate value for models to apprehend, and use the dual networks to quantify relations from shocks to liquidations and then to equilibrium prices. Although the exploration of ML models in the inverse demand function is still in its nascent stages, our framework pioneers a new direction in model explainability by facilitating a component-by-component understanding of the model.

We assess the performance of our model in two contexts. First, we evaluate it using simulated data with both linear and non-linear (i.e., exponential, arctangent) inverse demand functions. Our experimental results not only demonstrate the model's ability to accurately predict the final equilibrium asset prices, but also reveal a close alignment between predicted and true liquidations. These findings indicate that our framework can deliver comprehensible results without compromising learning performance. Second, we test our model on multiple assets, considering scenarios both with and without cross-impacts. While prior studies often assume that asset prices are only indirectly connected \cite{greenwood_vulnerable_2015, cont2019monitoring}, the experimental results highlight that our proposed approach can realistically capture price cross-impacts during financial contagion.

The remainder of this paper is organized as follows. In the next section, we provide an overview of related work on the inverse demand function in financial contagion and XAI methods in finance. We then delve into the details of our proposed dual network framework, including the objective definition and network design. Following this, we present our experimental setup, data simulation strategy, and results. We conclude the paper by pinpointing the potential limitations and proposing avenues for future research.

\section{Literature Review}
\subsection{Literature of Inverse Demand Function}
The contagion of losses through the financial system is often classified as either direct or indirect \cite{greenwood_vulnerable_2015, glasserman_how_2015}. Direct contagion -- also known as default contagion -- is characterized by the cascade of losses through the contractual obligations between financial institutions. In such an event, the failure of one institution can precipitate substantial losses for its direct counterparties; because of these losses, a counterparty may default on its own obligations propagating the initial shock. For default contagion, systemic risk is characterized by the counterparty network \cite{elsinger_risk_2006}.
In contrast, indirect contagion allows a crisis to spread through the financial system to firms that may not even have direct contractual relations with a distressed institution. In this work we focus on price-mediated contagion in which mark-to-market accounting triggers all institutions to write-down the value of their assets simultaneously; as asset values drop, firms may be forced to liquidate portions of their holdings leading to further impacts to asset prices which becomes self-reinforcing within the financial system \cite{shleifer_fire_2011}. As this indirect contagion need not follow the network of counterparties, indirect contagion effects can ripple across a wide range of seemingly unrelated banks and assets \cite{greenwood_vulnerable_2015}. Price-mediated contagion can cause orders of magnitude greater losses to the financial system than default contagion for these reasons, see e.g.,  \cite{greenwood_vulnerable_2015,calimani2022simulating, amicaosu2021fire}.
Hence, constructing models capable of capturing the impact of adverse shocks on asset liquidation, as well as their subsequent influence on market prices, holds pivotal importance. 

The relationship between trading activity and the resulting asset prices has, previously, been mathematically modeled through the use of ``inverse demand functions'' \cite{amini_uniqueness_2016,feinstein_effects_2017}. 
Within the price-mediated contagion literature, the two most prominent forms for the inverse demand function are \emph{linear} and \emph{exponential} functions. 
Within \cite{greenwood_vulnerable_2015}, a linear price impact model is assumed in which the sale of each asset share has a constant absolute impact on the price.
Alternatively, Cifuentes et al. \cite{cifuentes_liquidity_2005} utilize an exponential inverse demand function so that the sale of each asset share has a constant relative impact on the price.
Despite the easy interpretation of these example inverse demand functions, neither of these forms is consistent with financial practice. This problem was previously highlighted by \cite{bichuch_endogenous_2022} which presented a theory for the construction of these price impacts through the equilibrium of an exchange economy.

Herein we consider a data-driven approach to learning the inverse demand function. Specifically, our model leverages a deep neural network for the formulation of the inverse demand function, as opposed to adopting a pre-defined analytical form. This approach confers us the flexibility to encapsulate both linear and non-linear relations drawn from the empirical data.
Contrary to preceding studies, which postulate the price of assets are merely indirectly connected (see, e.g., \cite{greenwood_vulnerable_2015,cont2019monitoring}), our proposed approach is able to model realistic price cross-impacts.
However, this modeling exercise is hindered by data availability. Specifically, in practice, only the initial shocks and the final equilibrium asset prices are typically observable. The realized asset liquidations and portfolio rebalancing actions undertaken by the financial institutions are obscured by the financial system itself. As such, we propose a novel two-step approach in which: (i) the initial shocks map into predicted equilibrium ``liquiations'' and (ii) these values are fed in to the second neural network, which serves as an inverse demand function, to predict the resulting asset prices. As noted, the second neural network directly produces a data-driven equivalent of the inverse demand function even in the absence of observable liquidation data. Experimental results, as discussed in the subsequent sections, reveal a close alignment between the predicted liquidations from the first neural network and the true liquidations in simulated data.

\begin{figure*}[!t]
  \centering
  \includegraphics[width=0.9\linewidth]{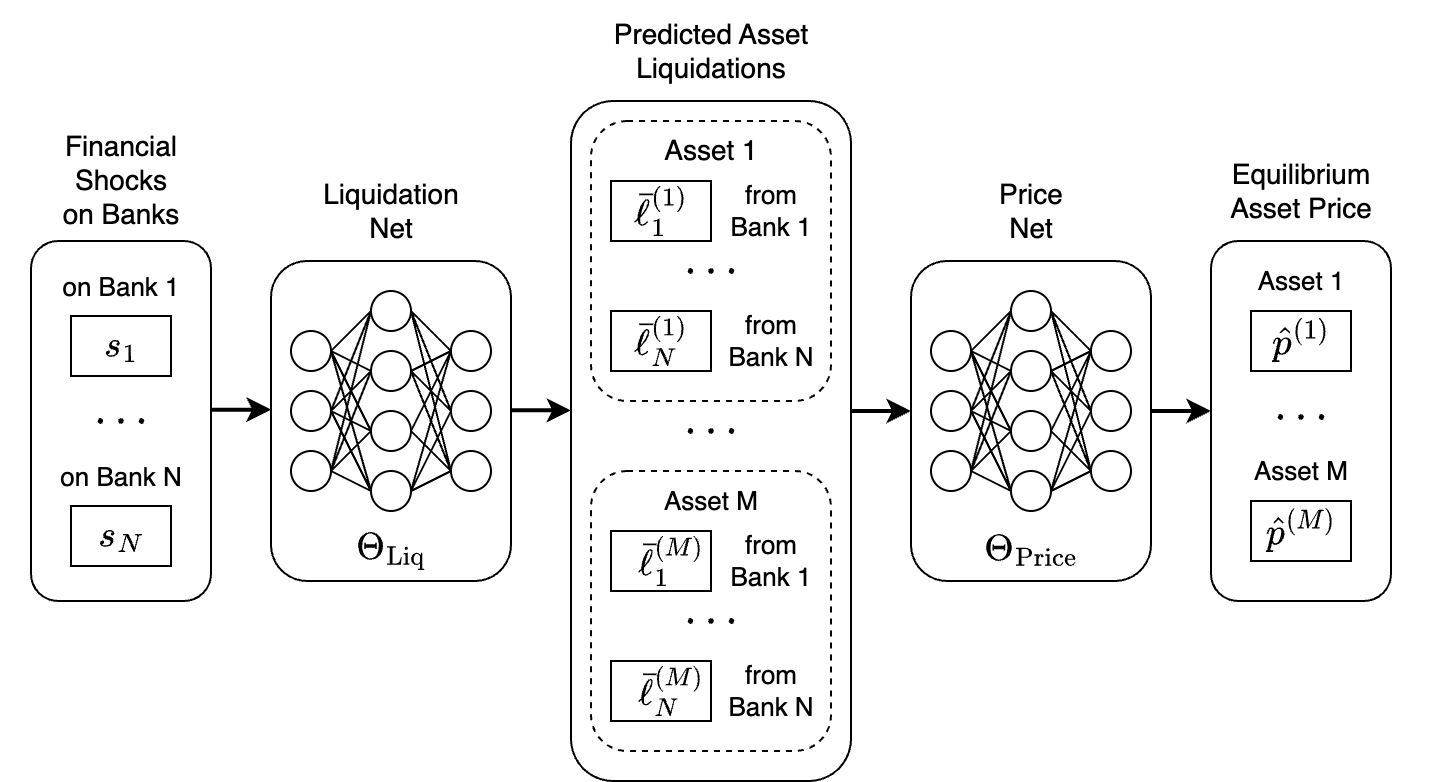}
  \caption{Framework Overview of the Dual Neural Networks}
  \label{fig: DNN}
\end{figure*}

\subsection{Literature of Explainable AI in FinTech}

Capitalizing on a wealth of data and advanced computational resources, the advent of ML -- specifically deep learning -- has brought forth a substantial transformation in the financial sector \cite{lecun_deep_2015}. The intersection of finance and technology, often termed as FinTech, has revolutionized the financial landscape by democratizing financial services, fortifying consumer protection, and refining risk management strategies \cite{gai_survey_2018,allen_survey_2020}. Academically, a growing body of literature also sheds light on the untapped potential of machine learning within the ambit of FinTech \cite{collins_artificial_2021,cao_ai_2022}.
However, the ML deployment in real-world financial applications poses significant challenges, chiefly due to their inherent lack of transparency and explainability \cite{misheva_explainable_2021}. Consider contemporary deep learning (DL) models as an example: Despite the superior accuracy when compared to traditional models like linear regression or decision trees, they are often viewed as ``black boxes'' that provide scant insightful information, which curtails their wider adoption within the finance sector \cite{bussmann_explainable_2020}. Transparency and explainability are critical factors in the development of trustworthy and reliable FinTech solutions for two essential reasons.

First, in highly-regulated domains like finance, regulators mandate rigorous transparency requirements for the implementation of advanced technologies. These measures aim to ensure traceability of decisions \cite{zheng_finbrain_2019}, adherence to legal regulations \cite{gomber_fintech_2018}, and observance of privacy standards \cite{weber_applications_2023}. With the massive amount of data generated in recent years, policymakers, legislators, and regulators all require explainable models to fulfill their expanding responsibilities and establish more proactive, data-driven regulation and surveillance approaches. Other stakeholders, including rating agencies and financial institutions, demand model transparency to evaluate their fairness in terms of industry or regional bias \cite{zhang_explainable_2022}. Even individual investors are wary of adopting models if the decision-making process remains opaque to them, regardless of the high accuracy these models might demonstrate.
Second, the ``black-box'' nature of models can expose them to potential vulnerabilities. Studies have shown that models may erroneously interpret similar yet distinct input features, leading to identical outputs. For example, computer vision researchers reveal that DL models struggle to differentiate between yellow-and-black stripe patterns of school buses and sticker-laden parking signs \cite{marcus_deep_2018}. Consequently, a lack of model interpretability and auditability could precipitate serious repercussions, potentially engendering macro-level risks that may cause unforeseen societal disruptions or harm \cite{goodman_european_2017}. Moreover, existing research indicates that attempts to enhance a model's explainability in finance often fail to elucidate the financial implications of the model in a manner comprehensible to average users. Instead, these efforts primarily serve machine learning engineers, assisting them in debugging or refining the models \cite{zhang_explainable_2022,bhatt_explainable_2020}. Hence, the pressing need for explainable AI in the finance field arises not only from the models' intrinsic requirement for robustness but also from the demands of diverse stakeholders and society at large.

To address these challenges, researchers propose the concept of XAI to enhance the transparency of ML models in FinTech.\footnote{We acknowledge that certain literature delineates the nuanced differences between interpretability and explainability in deep learning models. For instance, Misheva et al. \cite{misheva_explainable_2021} posit that interpretability pertains to understanding the relationships between cause and effect, while explainability concerns comprehending the functionality of each model component in human terms. However, as our work primarily focuses on modeling interpretable deep learning models rather than defining XAI, we employ these terms interchangeably throughout this paper as in \cite{du_techniques_2019}.}
XAI aims to generate more comprehensible models without compromising learning performance, thereby enabling humans to better understand, trust, and manage their artificially intelligent counterparts \cite{barredo_arrieta_explainable_2020}. Existing XAI methods can generally be categorized into two types: intrinsic interpretability and post-hoc interpretability \cite{du_techniques_2019}.

Intrinsic interpretability methods focus on constructing models based on human-oriented constructs, facilitating human comprehension of the transformation process from inputs to outputs. These methods rely on pre-programmed rules devised by human experts and are typically deterministic \cite{gomber_fintech_2018}, making it easier to identify potential biases and discriminatory practices within the algorithms \cite{dosilovic_explainable_2018}. However, these models' effectiveness is also constrained by their rigid, inflexible design. Such systems struggle to learn from new data or adapt to changing conditions, rendering them less effective in the fluid world of finance, which is charachterized by evolving markets and unpredictable economic conditions. Furthermore, even though some ML models offer high interpretability, their ease of interpretation can diminish as the scenario complexity escalates. For example, while decision trees are typically straightforward to explain, a real-world application of predicting mortgage defaults may require hundreds of large decision trees operating in parallel, which makes it challenging to intuitively summarize how the model functions \cite{bracke_machine_2019}.

Post-hoc interpretability methods, on the other hand, target the interpretation of complex ML models. As underscored by the trade-off between model complexity and performance \cite{du_techniques_2019,jin_pareto-based_2008}, advanced ML models often outperform decision trees or case-based reasoning models but, simultaneously, their complexity renders interpretability a challenge \cite{dosilovic_explainable_2018}. Post-hoc XAI studies mostly hinge on tools such as Local Interpretable Model-Agnostic Explanations (LIME) or Shapley Additive explanations (SHAP) \cite{ribeiro_why_2016,lundberg_unified_2017}. For instance, studies \cite{misheva_explainable_2021,bussmann_explainable_2020,ariza-garzon_explainability_2020} apply LIME and SHAP to explain credit scoring models, and their experimental results from 2.2 million peer-to-peer loan records indicated that both LIME and SHAP provided consistent explanations aligning with financial logic. To predict financial distress, Zhang et al. \cite{zhang_explainable_2022} propose an ensemble method paired with SHAP, while Park et al. \cite{park_explainability_2021} combine LIME with LightGBM- and XGBoost-based models to identify key features leading to bankruptcy. For asset management, Benhamou et al. \cite{benhamou_explainable_2021} integrate SHAP with gradient boosting decision trees to scrutinize a feature's impact from a set of 150 macroeconomic features during the 2020 financial meltdown.

In this study, we introduce a novel explainable framework that pre-defines learnable intermediate values for deep learning models to apprehend. 
Diverging from traditional XAI methods -- which typically explain the model post-training -- our framework imbues each model component with distinct financial significance \emph{a priori}. Although the exploration of ML for financial systemic risk is yet at a fledgling stage \cite{amini_optimal_2023}, our XAI paradigm pioneers a new direction in model explainability by facilitating a component-by-component understanding of the model. In the subsequent section, we will delve into the specifics of our model design.

\section{Method}

\subsection{Definition of the Objectives}

Our proposed explainable framework integrates two interconnected networks, with the output of the first network serving as a learnable intermediary value imbued with distinct financial significance. Here we apply our framework to model the inverse demand function, as it presents two pressing requirements that our framework is proficiently positioned to address.

First, as we noted previously, in typical real-world scenarios, only the initial financial shock and the ultimate equilibrium price are visible, while asset liquidation volumes remain elusive \cite{feinstein_effects_2017}. Our model bridges this gap by assigning the unobservable liquidation as the intermediate learning value, endowing this output with a distinct financial meaning. This approach not only addresses a critical shortcoming in existing XAI frameworks that lacks financial interpretability \cite{zhang_explainable_2022,bhatt_explainable_2020}, but also provides a versatile framework that can be expanded to broader contexts. Specifically, it is applicable in situations where only the input and output are apparent, yet we acknowledge the existence of certain indispensable hidden values within the process. Second, as a deep learning model, our framework excels in capturing the non-linearity between the shock, liquidation, and equilibrium price. While the prevailing literature commonly fits the relation as linear or exponential without justifying the persistence of such model types \cite{bichuch_endogenous_2022}, our framework fills this gap with a data-driven process and is readily scalable according to the data volume.

To this end, our proposed dual networks, purpose-built for the inverse demand function, fulfill two roles: (i) Deducing the liquidation volume for each asset sold by the respective banks, and (ii) Predicting the equilibrium price for each asset given financial shocks. In pursuit of these objectives, we denote the two interconnected neural networks within our framework as the Liquidation Net and the Price Net. As their names suggest, Liquidation Net ingests the shocks and produces the corresponding asset liquidation. Subsequently, Price Net employs the predicted liquidation as input to forecast the equilibrium asset price. In essence, the liquidation serves as an intermediary variable within the neural networks. The model is then trained using the loss derived from the discrepancy between the predicted and actual asset prices. Following the training phase, we examine the outputs of the Liquidation Net to infer liquidation values.

Figure~\ref{fig: DNN} presents a high-level illustration of the proposed architecture. We will elucidate the construction of the two networks, notations, as well as the detailed process, in the rest of this section.

\subsection{Model Architecture and Design}

We begin by outlining the transformation from financial shocks to asset liquidations via the Liquidation Net. 
Suppose there are $N$ banks and $M$ assets held by these banks in the price-mediated network. Accordingly, we denote banks by $n \in \{1,\ldots,N\}$, assets by $m \in \{1,\ldots,M\}$, and the impact of financial shocks on the banks by $\mathbf{s}=\left[s_{1}, s_{2}, \ldots, s_{N} \right]$.
Given the heterogeneity in bank sizes and asset portfolios, the assets held by banks are designated by $\mathbf{a}=\left[a^{(1)}_{1} , a^{(2)}_{1} , \ldots , a^{(M)}_{N} \right]$. For instance, $a^{(1)}_{1} = 1.8$ implies that bank 1 has 1.8 units of asset 1 in its possession.

We feed the financial shocks $\mathbf{s}$ into the Liquidation Net, a neural network made up of several fully connected layers. What differentiates our approach from traditional deep learning models is our embedding of the expected monotonic relationship between the shocks and liquidation \cite{amini_uniqueness_2016,bichuch_endogenous_2022}. For instance, it's logical that severe shocks should drive banks to sell off more assets, resulting in increased market liquidation. Therefore, larger shock values should correspond to more substantial liquidation outputs. To integrate this positive correlation, we tailor the Liquidation Net by imposing a clamp on the trainable parameters within each layer. Clamping ensures that the parameters remain non-negative (i.e., in the range $(0, \infty]$) during the training phase. We further adopt a Rectified Linear Unit (ReLU) activation function, defined as $\max\{0,x\}$, to guarantee that increased shocks lead to amplified asset liquidations. The inner workings of the Liquidation Net can be expressed mathematically as:
\begin{equation}
\bar{\boldsymbol{\ell}} = Liquid Net\left(\mathbf{s}\: ;\:  \Theta_{\text{Liq}}\right),
\end{equation}
where $\Theta_{\text{Liq}}$ represents the trainable parameters of the Liquidation Net and $\bar{\boldsymbol{\ell}}$ indicates the model's predicted liquidations. Considering that different banks will liquidate assets based on the shocks and their holdings, the predicted liquidation $\bar{\boldsymbol{\ell}}$ is comprised of assets' liquidations from each bank, represented as $\bar{\boldsymbol{\ell}}=\left[\bar\ell^{(1)}_{1}, \bar\ell^{(2)}_{1} , \ldots , \bar\ell^{(M)}_{N} \right]$ where $\bar\ell^{(m)}_{n}$ denotes the liquidation of asset $m$ by bank $n$.

Following asset divestiture by the banks, which generates market liquidation, we aggregate the liquidation values of assets as:
\begin{equation}
\hat{\boldsymbol{\ell}}=\left[\hat{\ell}^{(1)}, \hat{\ell}^{(2)}, \ldots, \hat{\ell}^{(M)} \right], \text{ where} \quad \hat{\ell}^{(m)} = \sum_{n = 1}^N a^{(m)}_{n} \bar\ell^{(m)}_{n}.
\end{equation}

Subsequently, these predicted liquidations of assets serve as inputs to the Price Net. Similar to the Liquidation Net, a monotonic relationship exists between liquidation and price, with higher liquidation values lead to lower prices. Hence, we also incorporate clamping and the ReLU function in the Price Net to maintain monotonicity. Contrary to the positive correlation in Liquidation Net, a negative correlation exists here. Thus, rather than directly outputting the predicted price, we output the negative predicted price as our final result. This process is depicted in the equation below:
\begin{equation}
\hat{\mathbf{p}} = Price Net\left(\hat{\boldsymbol{\ell}}\: ;\: \Theta_{\text{Price}} \right),
\end{equation}
where $\Theta_{\text{Price}}$ are the trainable parameters of the Price Net and $\hat{\mathbf{p}}=\left[\hat{p}^{(1)}, \hat{p}^{(2)}, \ldots, \hat{p}^{(M)} \right]$ denotes the predicted equilibrium prices of assets. The ground truth of the prices is represented as $\mathbf{p}$. The discrepancy between $\hat{\mathbf{p}}$ and $\mathbf{p}$ generates the loss, which is used to backpropagate and tune the trainable parameters (i.e., $\Theta_{\text{Liq}}$ and $\Theta_{\text{Price}}$). 
We summarize the notations in Table \ref{tab:notation_table} below. In the subsequent section, we perform experiments with the simulated dataset and evaluate our model's performance relative to the established benchmarks.

\begin{table}[!htbp]
\centering
\begin{tabular}{ll}
\toprule
 Notation & Description \\
\midrule
 $N$,$M$ & number of banks and assets, respectively \\
 $\mathbf{s}$ & financial shock impact on banks \\
 $\mathbf{a}$ & marketable asset holdings of banks \\
 $\mathbf{L}$ & liabilities of banks \\
 $\bar{\boldsymbol{\ell}}$ & predicted asset liquidations of banks  \\
 $\hat{\boldsymbol{\ell}}$ & predicted asset liquidations of assets \\
 $\hat{\mathbf{p}}$ & predicted equilibrium prices of assets  \\
 $\mathbf{p}$ & actual equilibrium price of assets  \\
 $\Theta_{\text{Liq}}$ & trainable parameters of Liquidation Net  \\
 $\Theta_{\text{Price}}$ & trainable parameters of Price Net  \\
\midrule
\end{tabular}
\caption{Summary of Used Notations}
\label{tab:notation_table}
\end{table}

\section{Experimental Results}
In this section, we undertake two detailed case studies to validate the performance of our proposed model. We first consider a single asset scenario with both linear and non-linear inverse demand functions; second, we consider a multi-asset scenario to investigate our model's capabilities to capture price cross-impacts and the utilized liquidation strategies.

\subsection{Case Study 1: Single-Asset Scenario}

Consider a system with $N=2$ banks and $M=1$ asset. We presume that the liabilities of the target banks, denoted by $\mathbf{L}$, are influenced by financial shocks. The magnitude of these liabilities varies according to the intensity of the financial shocks. Specifically, under the impact of these shocks, the liabilities randomly range between 0.6 and 0.85. We assume that both banks hold a single unit of the marketable illiquid asset, i.e., $a_{1}=a_{2}=1$. 

We study the performance of our dual network framework under linear, exponential, and arctangent inverse demand functions simulated with the contagion model from \cite{ feinstein_effects_2017,banerjee2020price}. \footnote{Consistent with these studies, we assume zero liquid and non-marketable assets throughout this paper.} We define the true linear inverse demand function as $\mathbf{p} = 1 - 0.15\ell$, the true exponential inverse demand function as $\mathbf{p} = \exp \left(-0.15 \ell\right)$, and the true arctangent inverse demand function as $\mathbf{p} = \frac{ \tan^{-1}(-\ell) + 2\pi}{2\pi}$. To evaluate the accuracy of these models, we compare the Mean Squared Error (MSE) of each model's predicted prices ($\hat{\mathbf{p}}$) against the actual asset prices ($\mathbf{p}$). It is noteworthy that none of the existing benchmarks can model the inverse demand function in the absence of liquidation data. Therefore, we validate our performance by comparing three modeling paradigms: 

\begin{itemize}
\item \textbf{Proposed Model}: Our proposed model involves a dual neural network structure, and only utilizes the financial shocks as input features. Initially, the shocks are used in the Liquidation Net to predict the liquidations of the target banks. Following this, the generated predictions are repurposed as inputs to the Price Net, thereby forming a sequential, interlinked predictive process.

\item \textbf{Linear Price Model}: This model substitutes the original Price Net with a linear regression model. The formulation of linear Price Net is based on the linear regression result between predicted liquidations and predicted asset prices.

\item \textbf{Inclusive Model}: This model integrates both the financial shocks and the true liquidations of target banks as observed inputs. Apart from the input modifications, the rest of this inclusive model's structure aligns with that of our proposed model.

\end{itemize}

Table \ref{tab:MSE} presents the comparison of MSE results for the predicted asset prices in the single-asset scenario. As summarized in this table, our model's accuracy closely aligns with the inclusive model. This is notable because the inclusive model benefits from additional input information (i.e., the actual bank liquidations). Also, our model's performance is consistent across both linear and non-linear inverse demand functions. 

\begin{table}[ht]
    \centering
    \setlength{\abovecaptionskip}{6pt} 
    \begin{tabular}{lSSS}
        \toprule
        Model Name & \multicolumn{3}{c}{Mean Squared Error (MSE)}\\
        \cmidrule(r){2-4}
        & {Linear} & {Exponential} & {Arctan} \\
        \midrule
        Proposed Model & \num{2.80e-7} & \num{6.80e-7} & \num{6.00e-8} \\
        Linear Price Model & \num{5.20e-7} & \num{8.50e-7} & \num{1.26e-5}  \\
        Inclusive Model & \num{9.00e-8} & \num{1.40e-7} & \num{4.00e-8} \\
        \bottomrule
    \end{tabular}
    \caption{MSE of the Predicted Asset Prices in the Single-Asset Scenario}
    \label{tab:MSE}
\end{table}

\begin{figure}[ht]
    \centering
    \begin{subfigure}[b]{0.33\textwidth}
        \includegraphics[width=\textwidth]{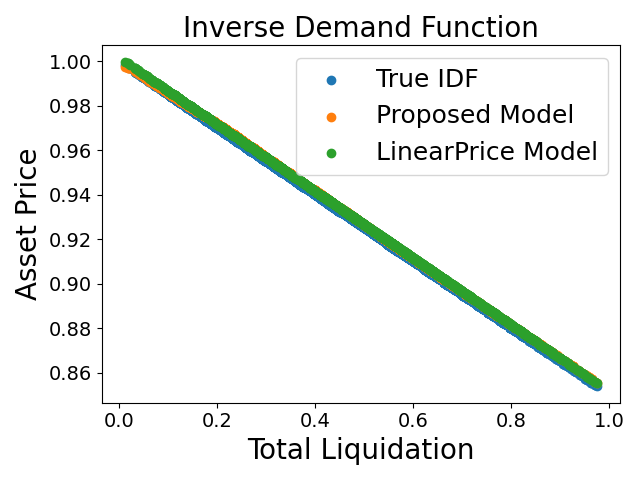}
        \caption{Linear}
        \label{fig:one_asset_inverse_demand_comparison_linear}
    \end{subfigure}
    \hfill
    \begin{subfigure}[b]{0.33\textwidth}
        \includegraphics[width=\textwidth]{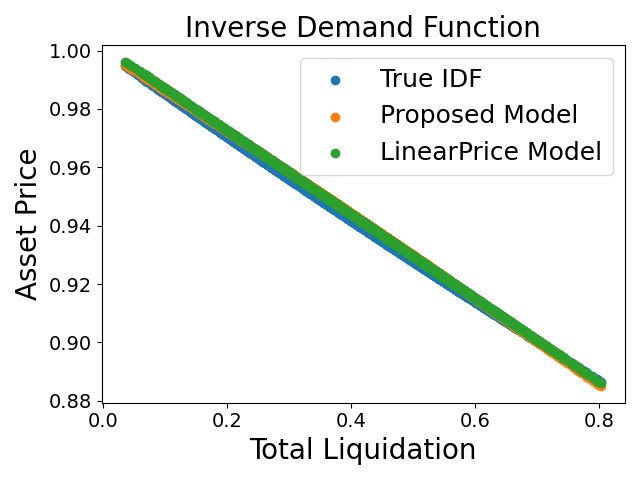}
        \caption{Exponential}
        \label{fig:one_asset_inverse_demand_comparison_exp}
    \end{subfigure}
    \hfill
    \begin{subfigure}[b]{0.33\textwidth}
        \includegraphics[width=\textwidth]{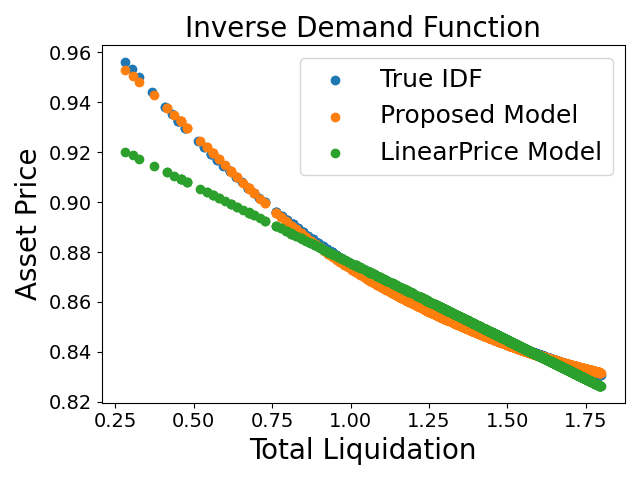}
        \caption{Arctangent}
        \label{fig:one_asset_inverse_demand_comparison_arctan}
    \end{subfigure}
    \caption{Predicted and Actual Inverse Demand Functions in the Single-Asset Scenario}
    \label{fig:IDF_linear}
\end{figure}

Furthermore, we find that, although the linear price model performs well when matched to a linear inverse demand function, it provides faulty results for the strongly non-linear arctangent inverse demand function. Thus, though it is tempting to utilize this simple structure, it can easily fail in practice. We note the high performance in the exponential case; as can be seen in Figure~\ref{fig:one_asset_inverse_demand_comparison_exp}, this inverse demand function is almost linear for the shocks considered.

In addition to validating the overall system performance (as shown in Table~\ref{tab:MSE}), we aim to investigate the capacity of our dual neural network structure to learn the hidden liquidation amounts. As indicated in Table~\ref{tab:model_correlations}, the liquidations ($\hat{\mathbf{\ell}}$) output by our proposed model exhibit an exceptionally high correlation with the actual liquidations ($\mathbf{\ell}$). Notably, this correlation remains robust across all three inverse demand function structures that we have employed.

\begin{table}[ht]
\centering
\setlength{\abovecaptionskip}{6pt} 
\begin{tabular}{l r}
\hline
\textbf{Dataset} & \textbf{Corr Coefficient} \\
\hline
Linear & 0.9999 \\
Exponential & 0.9995 \\
Arctangent & 0.9992 \\
\hline
\end{tabular}
\caption{Predicted and Actual Inverse Demand Functions in the Single-Asset Scenario}
\label{tab:model_correlations}
\end{table}

It is important to note that in our dual network model, the predicted liquidations are only able to capture monotonic trends in response to financial shocks, rather than absolute values. To rectify this and enhance the model's efficacy to predict absolute liquidation values, we introduce a linear scaling procedure. This process involves scaling the predictions to a range determined by the minimum and maximum financial shock values. The scaling process is succinctly described by the following formula:
\begin{equation}
\label{eq:scaling}
{ {\mathbf{\ell}^{*}} = (\hat{\mathbf{\ell}} - \hat{\mathbf{\ell}}_{\text{min}}) \times \frac{\mathbf{\ell}_{\text{max}} - \mathbf{\ell}_{\text{min}}}{\hat{\mathbf{\ell}}_{\text{max}} - \hat{\mathbf{\ell}}_{\text{min}}} + \mathbf{\ell}_{\text{min}}},
\end{equation}
where 
$\hat{\mathbf{\ell}}$ represents the outputs from the Liquidation Network, ${\mathbf{\ell}^{*}}$ represents the scaled predicted liquidations, and $\mathbf{\ell}$ represents the actual liquidations. This scaling procedure is applicable as, though the actual liquidations $\mathbf{\ell}$ are typically unobserved in practice, the response to minimal shocks {($\mathbf{\ell}_{min} \approx 0$)} or maximum shocks {($\mathbf{\ell}_{max} \approx a_1 + a_2$)} can be accurately predicted from balance sheet information. Following this, Equation (\ref{eq:scaling}) scales the initial, unscaled predictions of liquidations $\hat{\mathbf{\ell}}$ to ${\mathbf{\ell}^{*}}$, and ${\mathbf{\ell}^{*}}$ perfectly aligns with the actual liquidations after this scaling procedure is applied.

As we have successfully predicted the hidden bank liquidations and asset prices, we aim to illuminate the price impact between bank liquidations and equilibrium asset prices. This understanding is valuable for financial authorities in the construction of effective stress tests and regulatory policies. Figure \ref{fig:IDF_linear} provides an illustrative comparison between the predicted and actual inverse demand functions. The x-axis represents the total liquidation of assets, i.e., ${\mathbf{\ell}^{*}_{1}} + {\mathbf{\ell}^{*}_{2}} $, while the y-axis shows the corresponding asset price. It demonstrates a remarkable agreement between the predicted and actual inverse demand functions, underscoring the predictive accuracy of our model. Notably, this agreement persists across both linear and non-linear inverse demand functions. In fact, in comparing our proposed model to the linear price model, we find comparable performance for the linear and exponential inverse demand functions but significantly improved performance for the arctan inverse demand function.

\subsection{Case Study 2: Multi-Asset Scenario}



Consider now a financial system with $N = 2$ banks and $M = 2$ assets. 
We fix the total units of both assets to be 1, i.e., $a^{(k)}_1 + a^{(k)}_2 = 1$ for $k = 1,2$. However, in order to investigate a more complex system than in Case Study 1, we set $a^{(1)}_{1}=0.4,~a^{(2)}_{1}=0.6,~a^{(1)}_{2}=0.6$, and $a^{(2)}_{2}=0.4$. The banks' liabilities $\mathbf{L}$, fluctuate based on the intensity of the financial shocks, randomly spanning a range between 0.6 and 0.9. As discussed in the multi-asset framework of Section 4 in \cite{bichuch_endogenous_2022}, we consider the price cross-impacts that liquidating one asset can have on the price of another. This impacts, commonplace in real-world financial markets, stands in contrast to the typical, simplifying assumption that there are no cross-impacts as taken in previous studies (e.g., \cite{greenwood_vulnerable_2015, banerjee2020price}). We use this setting to simulate datasets with linear inverse demand functions that include or exclude price cross-impacts. 


Drawing upon the financial contagion model outlined by \cite{braouezec_strategic_2019,greenwood_vulnerable_2015,feinstein_effects_2017}, we validate the performance of our model under a proportional liquidation strategy. This strategy implies that each bank liquidates its portfolio proportionally to its holdings, enabling us to assess how our model learns banks' asset liquidation patterns, which can vary due to differences in portfolio holdings. Specifically, if a bank's portfolio comprises 60\% of Asset 1 and 40\% of Asset 2, then, when the bank liquidates, it disposes of 60\% of Asset 1 and 40\% of Asset 2.
We would like to emphasize that the proportional liquidation strategy has been extensively examined and validated in existing literature, as evidenced by works such as \cite{greenwood_vulnerable_2015,cont2019monitoring,banerjee2020price}.

\begin{table}[ht]
    \centering
    \setlength{\abovecaptionskip}{6pt} 
    \begin{tabular}{lSS}
        \toprule
        Model Name & \multicolumn{2}{c}{Mean Squared Error (MSE)}\\
        \cmidrule(r){2-3}
        & {Without Cross-Impacts} & {With Cross-Impacts} \\
        \midrule
        Proposed Model & \num{4.10e-7} & \num{8.00e-8} \\
        Inclusive Model & \num{1.70e-7}& \num{4.00e-8} \\
        \bottomrule
    \end{tabular}
    \caption{Sum of the MSEs of the Predicted Asset Prices in the Two-Asset Scenario}
    \label{tab:MSEtwoasset}
\end{table}

Based on the results presented in Table \ref{tab:MSEtwoasset}, it is clear that our proposed model performs impressively in a complex multi-asset scenario. While the MSE of the proposed model is slightly higher compared to the inclusive model, the errors are still comparable. This demonstrates that our proposed model can accurately predict asset prices even when dealing with a greater level of complexity. 

\begin{table}[h]
\centering
\setlength{\abovecaptionskip}{6pt} 
\begin{tabular}{l c c}
\hline
\textbf{Dataset} & \textbf{Corr Coefficient Asset 1} & \textbf{Corr Coefficient Asset 2} \\
\hline
Without Cross-Impacts & 0.9954 & 0.9523\\
With Cross-Impacts & 0.9999 & 0.9998\\
\hline
\end{tabular}
\caption{Correlation Coefficients between the Predicted and Actual Liquidations in the Two-Asset Scenario}
\label{tab:MSEtwoassetcorr}
\end{table}

Furthermore, as demonstrated in Table \ref{tab:MSEtwoassetcorr}, the predicted liquidations are extremely highly correlated to the true liquidations for both assets, either with or without cross-impacts. We wish to note that the same linear scaling procedure~\eqref{eq:scaling} can be adopted herein so as to improve the interpretability of our results.

Finally, we want to investigate our model's ability to capture the true inverse demand function with price cross-impacts. Herein, the true inverse demand function for asset 1 is given by $\mathbf{p}_1 = 1 - 0.15\ell_1 - 0.015\ell_2$. For this experiment, we fit a multiple linear regression with the predicted price of asset 1 ($\hat{p}_1$) as the dependent variable, and the scaled predicted liquidations of both assets (${\mathbf{\ell}^{*}}$) as independent variables. This regression analysis, as displayed in Table \ref{tab:linear_reg_stat_test}, has estimated coefficients of $-0.149$ and $-0.015$ for the scaled predicted liquidations of assets 1 and 2 respectively, and an intercept of approximately $1.000$.

\begin{table}[!h]
    \centering
    \setlength{\abovecaptionskip}{6pt} 
    \begin{tabular}{lcccc}
        \toprule
        & \multicolumn{2}{c}{Linear Regression Results} & \multicolumn{2}{c}{Statistical Test} \\
        \cmidrule(r){2-3} \cmidrule(r){4-5}
        Variable & Estimate & Standard Error & t-value & p-value \\
        \midrule
        Predicted Liquidation of Asset 1 $  $ & -0.149 & 0.010 & 0.082 & 0.935 \\
        Predicted Liquidation of Asset 2 $  $ & -0.015 & 0.007 & -0.062 & 0.951 \\
        Intercept $ $ & 1.000 & 0.004 & 0.110 & 0.912 \\
        \bottomrule
    \end{tabular}
    \caption{Linear Regression Estimates and Statistical Hypothesis Test Results for the Predicted Price of Asset 1. The Null Hypothesis for the Coefficients are: $\hat{a} = -0.15$, $\hat{b} = -0.015$ and for the Intercept, $\hat{c} = 1$.}
    \label{tab:linear_reg_stat_test}
\end{table}

We derive two major conclusions from the statistical tests performed. First, we fail to reject the null hypothesis that these coefficients are the true values (i.e., the coefficients are $-0.15$ and $-0.015$ for assets 1 and 2, respectively, and the intercept is $1.000$), with p-values of 0.935, 0.951, and 0.912 for these coefficients and intercept. This indicates that the model's estimates align closely with the true values in the inverse demand function. Second, when we test the null hypothesis that the coefficients are equal to zero, we reject this hypothesis as illustrated with extremely low p-values (<0.001, 0.032, and <0.001) in Table \ref{tab:linear_reg_stat_test_2}. 
This provides significant statistical evidence that the neural network accurately identifies the real price cross-impacts.
\begin{table}[!h]
    \centering
    \setlength{\abovecaptionskip}{6pt} 
    \begin{tabular}{lcccc}
        \toprule
        & \multicolumn{2}{c}{Linear Regression Results} & \multicolumn{2}{c}{Statistical Test} \\
        \cmidrule(r){2-3} \cmidrule(r){4-5}
        Variable & Estimate & Standard Error & t-value & p-value \\
        \midrule
        Predicted Liquidation of Asset 1 & -0.149 & 0.010 & -14.9 & *** \\
        Predicted Liquidation of Asset 2 & -0.015 & 0.007 & -2.14 & * \\
        Intercept & 1.000 & 0.004 & 250.0 & *** \\
        \bottomrule
    \end{tabular}
    \caption{Linear Regression Estimates and Statistical Hypothesis Test Results for the Predicted Price of Asset 1. The Null Hypothesis for the Coefficients are: $\hat{a} = 0$, $\hat{b} = 0$ and for the Intercept, $\hat{c} = 0$.}
    \label{tab:linear_reg_stat_test_2}
\end{table}

\section{Conclusion}
Within this work, we proposed a novel two-step neural network method to learn the inverse demand function from only partial information. Notably, although the liquidation values are unobserved, our proposed procedure is able to learn them with a high degree of accuracy. In this way, this procedure is designed so as to be \emph{ex ante} interpretable. To the best of our knowledge, this is the first study capable of modeling the inverse demand function in the absence of liquidation data. Given the robust performance observed in our numerical case studies, this method can be practically employed to construct a data-driven inverse demand function rather than the exogenous forms typically used in practice.

Our work further highlights a pair of promising avenues for future research. First, despite our model effectively capturing cross-impacts in multi-asset scenarios, our investigation focused on scenarios involving a limited number of assets and banks. Real-world cases, however, may encompass a more extensive range of assets or banks, often characterized by complex relational structures. Thus, future research can integrate more comprehensive, interconnected datasets to derive profound insights.

Second, we deliberately maintain a simple form of neural networks in our framework, in order to ensure a consistent number of trainable parameters with benchmark models \cite{rudin2019we}. Our experiments attest that the proposed framework achieves a comparable level of price prediction accuracy, even when partial information is employed. Considering the inherent noise of real-world data, the incorporation of more sophisticated deep learning architectures could further enhance the effectiveness of our proposed method. For example, replacing the fully connected networks with Graph Neural Networks (GNNs) could be a compelling development. Since GNNs consider both the features and relations of banks during predicting, this transition can offer a potential improvement to the model's adaptability in practical scenarios.

\bibliographystyle{unsrt}  
\bibliography{references}

\end{document}